\documentclass[11pt]{article}

\usepackage{amsmath,amssymb}
\usepackage{slashed}
\usepackage{xcolor} 
\usepackage{graphicx}
\usepackage{url}
\usepackage{cite}

\newcommand{\beq}{\begin{equation}}
\newcommand{\eeq}{\end{equation}}
\newcommand{\bea}{\begin{eqnarray}}
\newcommand{\eea}{\end{eqnarray}}
\newcommand{\nn}{\nonumber \\ }

\newcommand{\hc}{\mathrm{h.c.}}
\newcommand{\eps}{\epsilon}

\newcommand{\cO}{{\cal O}}
\newcommand{\cL}{{\cal L}}

\newcommand{\fref}[1]{Fig.~\ref{fig:#1}} 
\newcommand{\eref}[1]{Eq.~\eqref{eq:#1}} 

\newcommand{\sref}[1]{Section~\ref{sec:#1}}
\newcommand{\tref}[1]{Table~\ref{tab:#1}}

\begin{document}

\vspace*{-2cm}
\begin{flushright}
 LPT Orsay 14-23 \\
\vspace*{2mm}
\today
\end{flushright}

\begin{center}
\vspace*{15mm}

\vspace{1cm}
{\Large \bf 
Exotic Higgs decays in the golden channel 
} \\
\vspace{1cm}

{\bf Adam Falkowski and Roberto Vega-Morales}

 \vspace*{.5cm} 
Laboratoire de Physique Th\'eorique, CNRS -- UMR 8627, \\
Universit\'e de Paris-Sud 11, F-91405 Orsay Cedex, France
\vspace*{.2cm} 

\end{center}

\vspace*{10mm}
\begin{abstract}\noindent\normalsize.

The Higgs boson may have decay channels that are not predicted by the Standard Model.   
We discuss the prospects of probing exotic Higgs decays at the LHC using the 4-lepton final state. 
We study two specific scenarios, with new particles appearing in the  intermediate state of the $h \to 4 \ell$ decay. 
In one, Higgs decays to a Z boson and a new massive gauge boson, the so-called hidden photon. 
In the other, Higgs decays to an electron or a muon and a new vector-like fermion.  
We argue that the upcoming LHC run will be able to explore a new parameter space of these models that is allowed by current precision constraints.
Employing matrix element methods, we use the full information contained in the differential distribution of the 4-lepton final state to extract the signal of exotic decays.  
We find that, in some cases, the LHC can be sensitive to new physics even when the correction to the total  $h \to 4 \ell$ rate is of the order of a percent. 
In particular, for the simplest realization of the hidden photon  with the mass between 15~and~65~GeV, new parameter space can be explored in the LHC run-II. 

\end{abstract}

\vspace*{3mm}

\section{Introduction} \label{sec:intro}

The particle with mass $m_h \approx 125.6$~GeV discovered at the LHC is so far perfectly compatible with being the Standard Model (SM) Higgs boson \cite{Aad:2012tfa,Chatrchyan:2012ufa}. 
It is nevertheless conceivable that more in-depth studies will reveal its non-standard properties. 
In particular, the Higgs may have exotic decay channels, that is channels not predicted in the SM or predicted to occur with a negligible branching fraction. 
Many scenarios beyond the SM predict new Higgs decay channels, especially in the presence of  new  degrees of freedom with $m \lesssim m_h$.  
The existing LHC searches for exotic Higgs decays cover decays to invisible particles \cite{Chatrchyan:2014tja,Aad:2014iia}, to 4~photons \cite{ATLAS:2012soa} or 4~muons via new \cite{CMS:2013lea,Aad:2012kw} intermediate bosons,  to electron jets \cite{Aad:2013yqp}, and to long-lived neutral particles \cite{ATLAS:2012av,CMS:2012rza}. 
However many more interesting final states and topologies exist \cite{Blankenburg:2012ex,Harnik:2012pb,Isidori:2013cla,Huang:2013ima,Gonzalez-Alonso:2014rla,Curtin:2013fra}; see Ref~\cite{Curtin:2013fra} for a comprehensive review. 
It should be noted that the current Higgs data can easily accommodate an order $20$\% branching fraction for exotic decays, 
and even more if the Higgs production cross section is enhanced, and/or Higgs couplings to the SM matter are modified,  see \fref{brexotic}.   
Furthermore, the sizable Higgs production cross section  at the LHC  allows us to probe much smaller branching fractions: down to  $\sim 10^{-5}$ currently, and down to $\sim 10^{-9}$ in the future 100 TeV collider, as long as the final state is experimentally clean. 
All this makes exotic Higgs decays an attractive direction to search for new physics. 

One very promising \cite{Curtin:2013fra,Isidori:2013cla} signature for this kind of searches   is the so-called {\em golden channel}:  the $4 \ell$ final state, $\ell = e,\mu$, with two opposite-sign same-flavor lepton pairs.   
Thanks to the fully reconstructible kinematics, low background, and small systematic errors it was one of the early Higgs discovery channels despite the small branching fraction. 
At the same time, order one new physics corrections to the SM rate in this channel can be accommodated at this point. 
Assuming the Higgs production cross section is unchanged from the SM, the event rates reported in  Refs.~\cite{ATLAS:2013nma,Chatrchyan:2013mxa} yield the 95\% CL limits on the additional partial decay widths: 
 \beq
 \label{eq:gammalimits}
{\Delta \Gamma_{h \to 4 \mu} \over \Gamma_{h \to 4 \mu}^{\rm SM}} <  0.90,  
\quad  
{\Delta \Gamma_{h \to 2 e 2\mu} \over \Gamma_{h \to 2 e 2\mu}^{\rm SM}} < 0.83,  
\quad    
{\Delta \Gamma_{h \to 4 e} \over \Gamma_{h \to 4 e}^{\rm SM}} < 1.27. 
\eeq 
For new physics contributing to all sub-channels the limit is  
\beq  
 \label{eq:gammalimits2}
{\Delta \Gamma_{h \to 4 \ell} \over \Gamma_{h \to 4 \ell}^{\rm SM}} < 0.52.    
\eeq 
Strictly speaking, the widths in \eref{gammalimits} and \eref{gammalimits2} should be weighted by the efficiency to experimental cuts, which may differ in  the presence new physics. 

\begin{figure}[h]
    \begin{center}
         \includegraphics[width=0.6\textwidth]{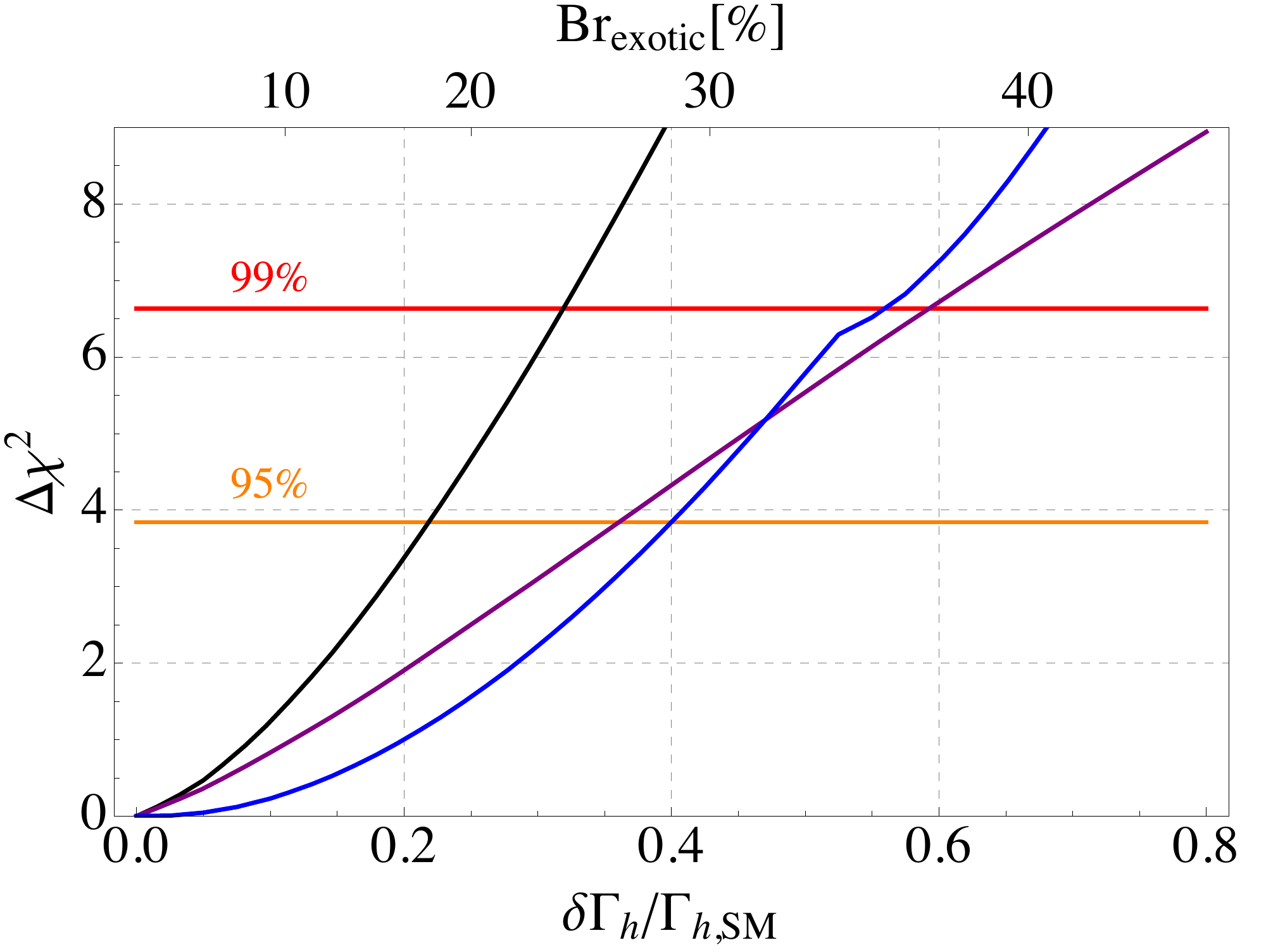}
\vspace*{-2mm}
          \caption{\footnotesize 
  Global fit to the Higgs  data in the presence of an exotic contribution to the Higgs decay width $\delta \Gamma_h$.
  The black curve assumes the Higgs production cross section and relative branching fraction to the SM matter are fixed at the SM values, which leads to  the indirect  limit  ${\rm Br}(h \to {\rm exotic}) \lesssim 18\%$ at 95\% CL.    
  This limit takes into account the uncertainty on the SM prediction of the gluon-fusion production cross-section which we take  as $14.7$\% \cite{Heinemeyer:2013tqa}. 
Leaving as a free parameter in the fit the gluon fusion production cross section (purple curve), and/or the Higgs branching fraction to b-quarks (blue curve), the limit is relaxed to  ${\rm Br}(h \to {\rm exotic}) \lesssim 30\%$. 
  If all effective Higgs couplings to the SM are left free  then only the model independent bound ${\rm Br}(h \to {\rm exotic}) \lesssim 80\%$ applies, based on the direct Higgs width measurement in CMS \cite{CMS:2014ala}. 
   }
\label{fig:brexotic}
\end{center}
\vspace*{-3mm}
\end{figure}

Apart from the event rate, the $4\ell$ final state offers far more information in the form of the differential distribution in the decay angles and lepton pair invariant masses.  
In this paper we investigate the possibility of using this information  to further constrain exotic decays of the Higgs boson. 
We employ the matrix element methods originally developed for the purpose of determining the structure of the Higgs couplings to the SM gauge bosons  \cite{Chen:2013ejz,Chen:2014pia,Chen:2014gka}. 
The starting point for our analysis is an analytic expression for the fully differential $h\rightarrow 4 \ell$ matrix element, with and without the new physics contribution. 
Using this matrix element, we construct a likelihood function for a data set containing a number $N$ of 4-lepton events. 
This likelihood function is then used to estimate the statistical significance for discrimination  between the SM and exotic decays hypotheses as a function of $N$.  

We study two simple models that can accommodate sizable exotic branching fractions in the golden channel without violating current experimental constraints. 
The first one contains a new light gauge boson $X$ coupled to the SM via the hypercharge portal  $\epsilon X_{\mu \nu} B_{\mu \nu}$  \cite{Holdom:1985ag}. 
The kinetic mixing induces the coupling of $X$ to the electromagnetic current, and also the mixing between the Z boson and $X$. 
As a result, the Higgs boson can decay as  $h \to X Z$  when it is kinematically allowed. 
When both $X$ and $Z$ decay leptonically, this new Higgs decay mode contributes to the $4 \ell$ final state. 
Another model we study here contains a new heavy vector-like charged lepton $E$ transforming as $(1,1)_{-1}$ under the SM gauge group.  
After electroweak symmetry breaking $E$ mixes with one of the SM leptons via Yukawa couplings.
As a result, one obtains non-diagonal couplings  to the Z and Higgs  boson of the form  $ Z_\mu \bar E_L \gamma_\mu \ell_L + \hc$ and $h \bar E_R \ell_L h + \hc$. 
These couplings mediate the $h \to E \ell \to Z \ell \ell$ cascade decay that,  for leptonic Z decays,  again contributes to the 4-lepton final state. 
 
The paper is organized as follows. 
In \sref{models} we describe our models in more detail. 
In \sref{meth} we review the matrix element methods to extract information from the golden channel. 
Our results regarding the sensitivity of the golden channel to exotic Higgs decays  are contained in \sref{results}.

\section{Models} \label{sec:models}

In this section we study two scenarios where new light degrees of freedom can modify Higgs decays in the golden channel.  
One has a new light vector field (the hidden photon) kinetically mixing with the SM hypercharge. 
The other has a new vector-like fermion with quantum numbers of the SM right-handed electron that mixes via a Yukawa coupling with one of the SM charged leptons.  
We determine the region of the parameter space of these models allowed  by precision measurements, and we discuss the limits on the  branching fraction for exotic Higgs decays imposed  by these constraints.  

\subsection{Hidden Photon}

The first model we study has  cascade decay $h \to Z X \to 4 \ell$ mediated by a new neutral vector boson. 
Consider a massive abelian gauge field  $X_\mu$ interacting with the SM only via the hypercharge portal:  
\beq
\cL = \cL_{\rm SM}  - { 1 - \eps^2 \cos^{-2}\theta_W \over 4} \hat X_{\mu \nu} \hat X_{\mu \nu}
 + {1 \over 2 } \hat m_X^2 \hat X_\mu \hat X_\mu +   {\eps \over 2 \cos \theta_W} B_{\mu \nu} \hat X_{\mu \nu}. 
\eeq 
Here $\theta_W$ is the Weinberg angle, and the non-standard normalization of the $X$ kinetic term is introduced for future convenience. 
We assume $\eps \ll 1$ and determine the spectrum and couplings perturbatively in $\eps$.  
The mass term $\hat m_X$ could be generated  via the St\"uckelberg mechanism, or via an expectation value  of a hidden sector Higgs field; 
in the latter case we will assume the corresponding hidden Higgs boson is heavy enough such that it does not affect the hidden photon decays.  
We are interested in $\hat m_X  \ll m_Z$, such that $X$ can have a non-negligible effect on Higgs decays. 

To work out the model's phenomenology it is convenient to remove the kinetic mixing by redefining  the hypercharge gauge field: 
$B_\mu \to B_\mu +  \cos \theta_W^{-1} \hat X_\mu$.     
The kinetic terms are now diagonal and canonically normalized, but after the EW breaking  the $Z$ and $X$ bosons mix via the mass terms, 
\beq
\cL_{\rm mass}  = {1 \over 2} \hat m_Z^2 \hat Z_\mu \hat Z_\mu +   {1 \over 2} \left ( \hat m_X^2 + \eps^2 \hat m_Z^2 \tan^2\theta_W \right) \hat X_\mu \hat X_\mu 
 - \hat m_Z^2 \eps \tan\theta_W  \hat X_\mu \hat Z_\mu,  
\eeq 
where $\hat m_Z = \sqrt{g_L^2 + g_Y^2} v/2$ and we denote $g_L$, $g_Y$ the SM gauge couplings of $SU(2)_L\times U(1)_Y$.  
To diagonalize the mass matrix  we need  the rotation 
\beq
\hat Z_\mu = \cos \alpha  Z_\mu  +\sin \alpha  X_\mu, 
\qquad 
\hat X_\mu = -\sin \alpha  Z_\mu  +\cos \alpha  X_\mu, 
\qquad
\alpha \approx \eps \tan \theta_W {m_Z^2 \over m_Z^2 - m_X^2}  + \cO(\eps^2). 
\eeq 
Mixing between the Z and exotic bosons is constrained electroweak precision observables.
In particular, it affects the mass of the Z boson,   
\beq
m_Z^2 = \hat m_Z^2 + \eps^2 {\tan^2 \theta_W \hat m_Z^4 \over m_Z^2 - \hat m_X^2} + \cO(\eps^3), 
\eeq 
and the Z boson couplings to matter,  
\beq
g_{Z,f} = \hat g_{Z,f} \left (1- \eps^2 {\tan^2 \theta_W  m_Z^4 \over  (m_Z^2 - m_X^2)^2} \right  ) 
- \eps^2 \sqrt{g_L^2 + g_Y^2} {\tan^2 \theta_W m_Z^2 \over  m_Z^2 -  m_X^2} Y_f ,   
\eeq 
where $\hat g_{Z,f} = \sqrt {g_L^2 + g_Y^2} (T^3_f - \sin^2\theta_W Q_f)$ is the Z boson coupling in the SM. 
Using the constraints from LEP-1 and SLC \cite{ALEPH:2005ab} and W mass \cite{Group:2012gb}  measurements for $m_X \ll m_Z$ we find 
\beq
|\eps| \lesssim 0.024 \ \sqrt{1- {m_X^2 \over m_Z^2}} \qquad {\rm at} \ 95\% \ {\rm C.L.}, 
\eeq
in agreement with Ref.~\cite{Hook:2010tw}. 
For $m_X$ below $9.3$~GeV one gets a stronger limit $|\eps| \lesssim 10^{-3}$  \cite{Bjorken:2009mm,Curtin:2013fra} based on $\Upsilon(2S,3S) \to \gamma \mu^+ \mu^-$ searches in  BaBar  \cite{Aubert:2009cp}. 

We turn to the couplings of the hidden photon. The couplings to the SM fermion are 
\beq
g_{X,f} = \epsilon \, e \left [ Q_f \left (1 - {\tan^2 \theta_W  m_X^2 \over m_Z^2 - m_X^2}  \right )
+ T^3_f  {m_X^2 \over \cos^2\theta_W (m_Z^2 - m_X^2)} \right ]. 
\eeq 
The new vector field couples to the electromagnetic current up to $\cO(m_X^2/m_Z^2)$ corrections,  hence the name {\em hidden  photon}. 
Assuming there's no other decay channels of $X$ (in particular, there is no decay to other particles in the hidden sector), for $m_X \ll m_Z$ one finds ${\rm Br} (X \to l^+ l^-) \approx 0.15$,  ${\rm Br} (X \to {\rm had}) \approx 0.55$,  while ${\rm Br} (X \to \nu \nu)$ is negligible.  Due to the mixing with $Z$, the hidden photon also acquires the coupling to the Higgs boson: 
\beq
\label{eq:hzx}
\cL_{hZX} =  c_{hZX} {m_Z^2  \over v} h Z_\mu X_\mu,  
\qquad    c_{hZX}  = {2 \eps \tan\theta_W m_X^2 \over m_Z^2 - m_X^2}  + \cO(\eps^2). 
\eeq 
Thus, all elements are in place for new contributions to the golden channel via the cascade decay $h \to Z X \to 4 \ell $. 
However, the coupling in \eref{hzx} is suppressed not only by $\eps$ but also by $m_X^2/m_Z^2$. 
For this reason,  the maximum ${\rm Br} (h \to Z X)$ does not exceed  $2.5 \times 10^{-4}$, as can be read off from  the right panel of \fref{brhxz}.
Currently, such a small branching fraction is not constrained by the observed $h \to 4 \ell$ event rate. 
Even scaling the present sensitivity to 300~fb${}^{-1}$ of data at 14 TeV LHC, the rate information alone does not allow one to explore the parameter space that is not excluded by precision measurements, see the left panel of \fref{brhxz}. 
Somewhat  stronger limits can be obtained when the input from the dilepton invariant mass distribution is used \cite{Curtin:2013fra}, but these limits are still weaker than the ones from electroweak precision tests.
In \sref{results} we will argue that the sensitivity can be further enhanced by using the full information contained in the differential distribution of $h \to 4 \ell$ decays. 
  
\begin{figure}[!h]
    \begin{center}
         \includegraphics[width=0.4\textwidth]{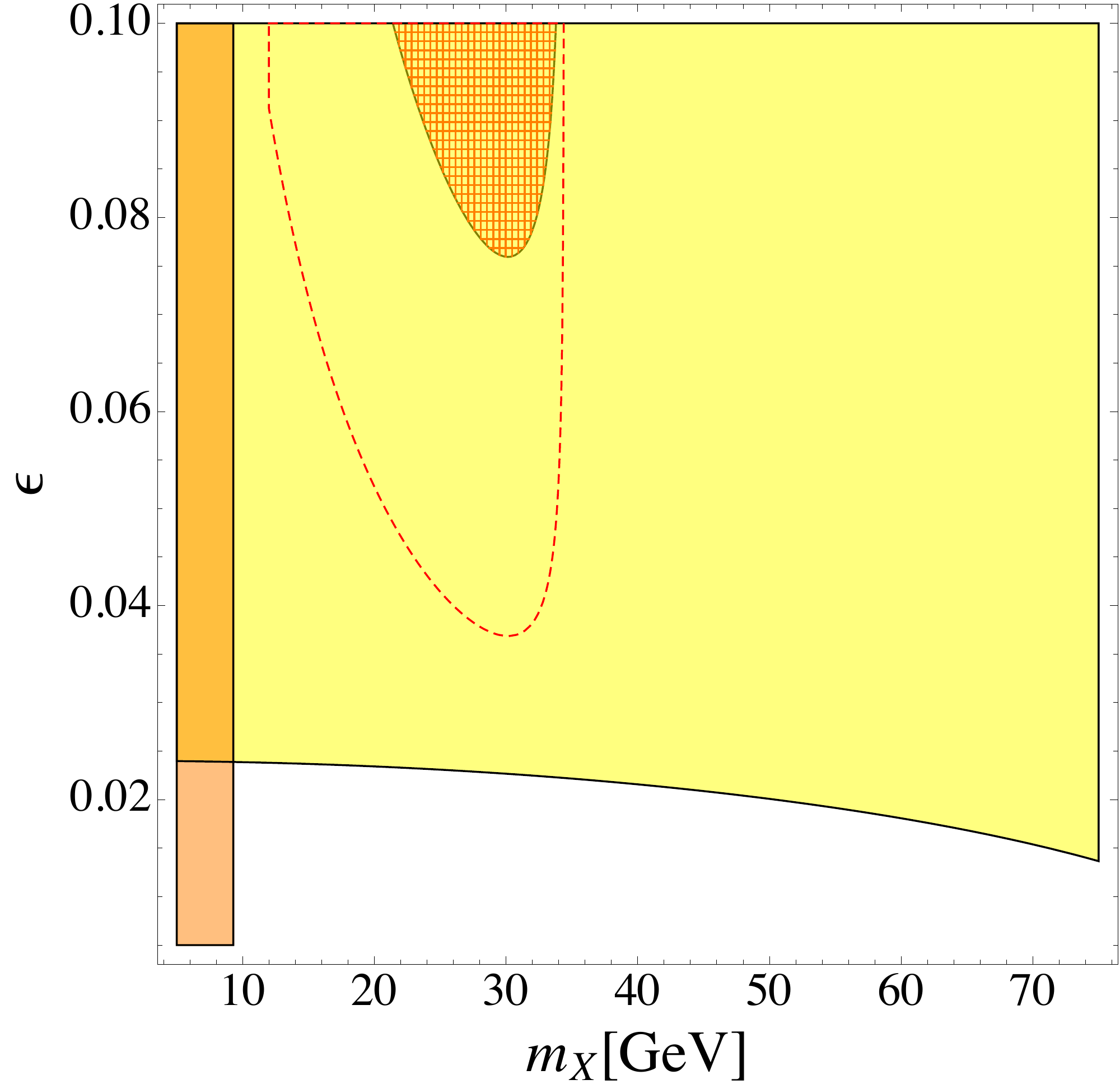}
     \quad         
      \includegraphics[width=0.55\textwidth]{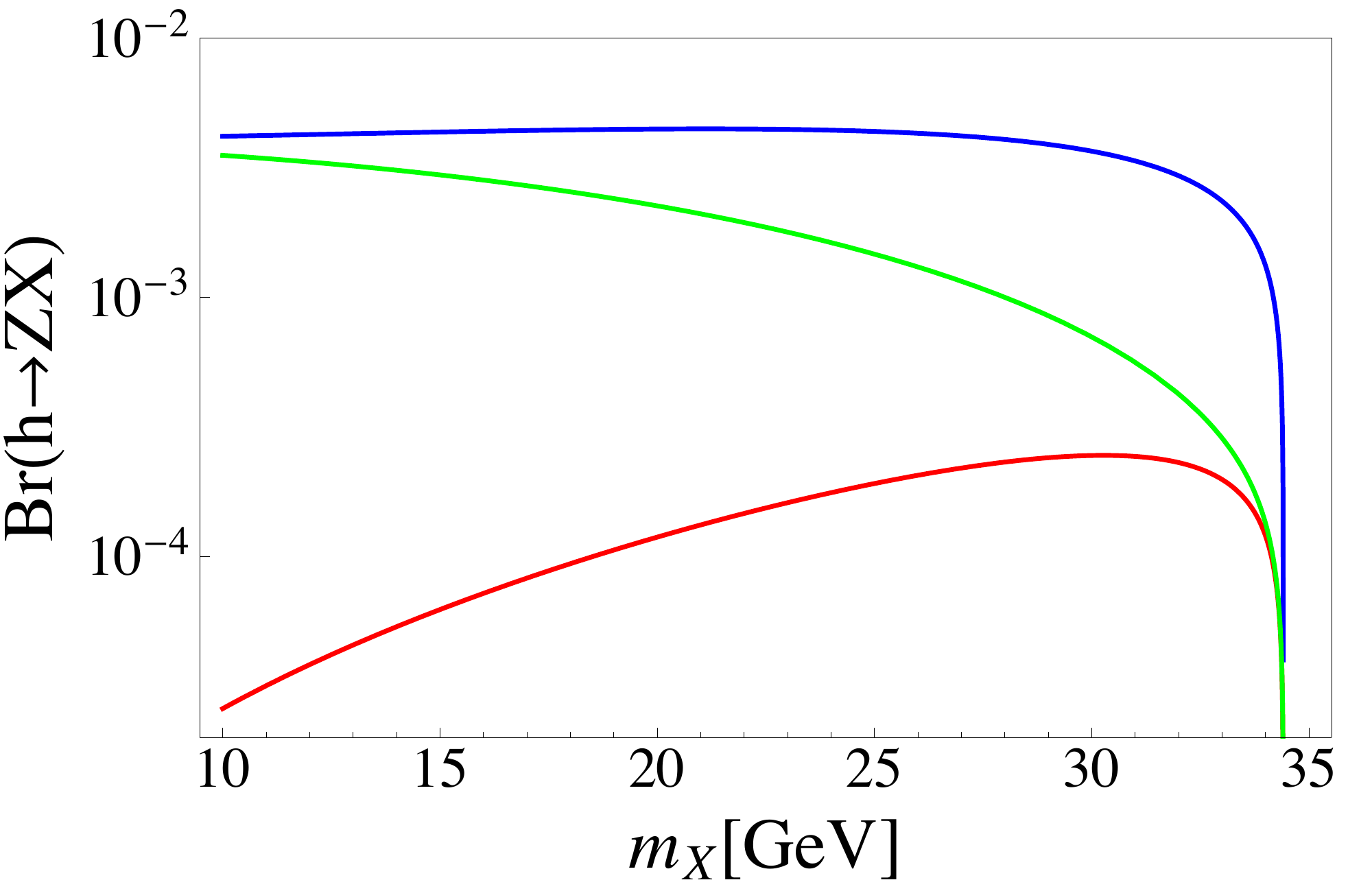}
\vspace*{-2mm}
          \caption{\footnotesize 
   {\bf Left:}  the parameter space in the mass vs.~mixing  plane for a hidden photon mixing with the SM hypercharge gauge boson. 
  For  this plot we assume $\eps_2= \eps _3 = 0$.  
   The yellow and orange areas are excluded respectively by direct BaBar searches and by electroweak precision constraints.
       The red mesh area is excluded by the observed $h \to 4 \ell$ event rate, taking into account  $h \to X Z$ decays with both $X$ and $Z$ on-shell, and  assuming the Higgs couplings to the SM matter are not modified).       
 The red dashed line shows an  estimated expected limit based on the 4-lepton event rate information with  300~fb${}^{-1}$ at 14~TeV LHC.                
       {\bf  Right:} The branching fraction for $h \to X Z$ in the hidden photon model for $\eps=0.02$ and  $\eps_2= \eps _3 = 0$ (red), $\eps_2=0.02$, $\eps _3 =0$ (blue),  and $\eps_2=0$, $\eps _3 =0.02$ (green).  
   }
\label{fig:brhxz}
\end{center}
\vspace*{-3mm}
\end{figure}

A larger 4-lepton branching fraction can be obtained by modifying the model. 
One way is to introduce mixing between the SM and the hidden Higgs boson $S$ that subsequently decays as $S \to XX$  \cite{Gopalakrishna:2008dv}. 
Here we consider another simple modification.   
One can introduce additional couplings between the hidden photon and the SM sector \cite{Davoudiasl:2013aya}: 
\beq
\label{eq:steroid}
\Delta \cL =  {\eps_2  \over  \cos \theta_W}   \left ( {|H|^2  \over v^2} - {1 \over 2} \right ) B_{\mu \nu} \hat X_{\mu \nu} 
+ {\eps_3 \over  \cos \theta_W}    {|H|^2  \over v^2}  \tilde B_{\mu \nu} \hat X_{\mu \nu} , 
\eeq   
where $\tilde B_{\mu \nu} = \eps_{\mu\nu\rho\sigma} \partial_\rho B_\sigma$. 
The new terms in $\Delta \cL$ induce new couplings of the Higgs boson to the Z boson and the hidden photon: 
\beq
\label{eq:dhzx}
\Delta \cL_{hXZ} =  - {h \over v}  \tan \theta_W \left ( \eps_2  X_{\mu \nu}  Z_{\mu \nu}  + \eps_3  X_{\mu \nu}  \tilde Z_{\mu \nu}     \right ) + \cO(\eps^2). 
\eeq 
In principle, the parameters $\eps_2$ and $\eps_3$ are not constrained by precision observables (although $|\eps_2| \gg |\eps|$ would be fine-tuning).\footnote{Note that the CP-odd kinetic mixing term  $\tilde B_{\mu \nu} \hat X_{\mu \nu}$ is a total derivative and has no physical consequences.}
Furthermore, the Higgs  couplings in \eref{dhzx} are not suppressed by  $m_X^2/m_Z^2$, unlike in the vanilla model. 
For these reasons,  this deformation of the hidden photon  model  allows for a sizable branching fraction  for $h \to X Z$ decay. 
In fact, the strongest constraints on $\eps_2$ and $\eps_3$ currently come from the $h \to 4 \ell$ searches.   
 
We note that for $\eps_{2,3} \neq 0$ the model also contains the $h X \gamma$ couplings:  
\beq
\label{eq:dhgx}
\Delta \cL_{hX\gamma} =   {h \over v}  \left ( \eps_2  X_{\mu \nu}  A_{\mu \nu}  + \eps_3  X_{\mu \nu}  \tilde A_{\mu \nu}     \right ) + \cO(\eps^2). 
\eeq 
It leads to an additional contribution to the $h \to 4\ell$ decay, with an off-shell photon  instead of Z. 
The size of this contribution strongly depends on the experimental cuts  on the final state leptons.\footnote{The inclusive $h \to 4l$ rate is IR divergent at the tree-level when diagrams with an intermediate photon are included.} 
We find that for the standard CMS cuts  the photon mediated contribution affects the new physics corrections to the $4\ell$ event rate by an ${\cal O}(1)$ factor.  
Another consequence of the couplings in \eref{dhgx} is the presence of  $h \to X \gamma$ decays with an off-shell photon. 
The branching fraction is larger than that for $h \to X Z $ decays because the $h X \gamma$ coupling is larger by $\tan^{-1} \theta_W$,  and because there is less  phase space suppression. 
For example, for $\eps_{2} = 0.02$ or  $\eps_{3} = 0.02$ one finds ${\rm Br}(h \to X \gamma) \approx 10$~\%. 
Therefore this version of the hidden photon model can also be probed in the $h \to \ell^+ \ell^- \gamma$ final state. 
We postpone to a future publication  quantitative studies of the sensitivity of  the $h \to \ell^+ \ell^- \gamma$ channel to exotic Higgs decays.

\subsection{Vector-like  Lepton}

The other scenario we study in this paper is the one where Higgs decays can proceed as $h \to E  l  \to  Z \ell^+ \ell^- \to 4 \ell $, mediated by a new charged lepton mixing with the SM leptons.  
Consider the SM  extended by a vector-like fermion $E$ transforming under the SM gauge group as $(1,1)_{-1}$, thus having quantum numbers of the right-handed electron.  
We assume $E$ mixes with one of the  SM  charged leptons via Yukawa couplings. 
The part of the Lagrangian giving rise to the vector-like and SM lepton masses is given by  
\beq
\cL =  - y \bar \ell_R H^\dagger l_L   - M_E \bar E_R E_L       - Y \bar E_R H^\dagger l  +\hc ,   
\eeq   
where $l_L = (\nu_L,\ell_L)$, and $\ell$ could be electron, muon, or tau.  
The first term is the usual SM lepton Yukawa coupling.  The second is a  vector-like mass $M_E$ of the heavy fermion. 
The last term leads to a mixing between the vector-like and the SM lepton after electroweak  symmetry breaking. 
We assume $Y v \ll M_E$ and $y v \ll M_E$, in which case the lepton mass eigenstates of the mass matrix can be worked out perturbatively  in $v$. 
To diagonalize the mass matrix we make the rotation
\bea
\ell_L \to \cos \alpha_L \ell_L  + \sin \alpha_L E_L,   & \qquad &  E_L \to -\sin  \alpha_L \ell_L  + \cos \alpha_L E_L,
 \nn
\ell_R \to \cos \alpha_R \ell_R  + \sin \alpha_R E_R,  & \qquad &   E_R \to -\sin  \alpha_R \ell_R  + \cos \alpha_R E_R,
\eea 
where the mixing angles are 
\beq
\alpha_L =   {Y v  \over \sqrt 2 M_E} \left (1  + \cO(v^2/M_E^2) \right ) , 
\qquad 
\alpha_R =  \cO(v^2/M_E^2). 
\eeq
Thus, at the leading order, only left-handed charged leptons mix with the vector-like lepton. 
The mass  of the heavy lepton is approximately $M_E$, and the mass of the SM lepton is approximately $y v /\sqrt 2$, up to $  \cO(v^2/M_E^2)$ corrections. 

Because $E_L$ and $\ell_L$ have different quantum numbers under the EW group, the mixing affects the lepton couplings to W and Z. 
At the leading order one obtains non-diagonal lepton couplings to W and Z bosons, 
\beq
 \cL = {g_L \over \sqrt 2} \alpha_L W_\mu^+ \bar \nu _L \gamma_\mu E_L  - {\sqrt{g_L^2 + g_Y^2} \over 2} \alpha_L  Z_\mu \bar \ell_L \gamma_\mu E_L    
\eeq  
These couplings allow the heavy lepton to decay as $E \to Z \ell$ or as  $E \to W \nu$, and we assume here that $E$ has no other decay channels. 
For $M_E$ close to $m_Z$  the branching fractions  strongly depend on $M_E$ (due to the phase space suppression), and ${\rm Br} (E \to Z \ell)$ varies between 10\% and 25\% for  $M_E$ between 100 and 125~GeV.  
The Higgs boson also obtains non-diagonal couplings to the leptons: 
\beq
\cL = - {Y  \over \sqrt 2} h  \bar E_R \ell_L + \hc  . 
\eeq
At the end of the day, for $m_Z < M_E < m_h$, the  Higgs boson can cascade decay as $h \to E  l  \to  Z \ell^+ \ell^- \to 4 \ell $. 

The mass of the heavy lepton is constrained by direct LEP-2 searches  $M_E \gtrsim 103$~GeV \cite{Beringer:1900zz}.    
So far the LHC experiments have not  provided new limits on $M_E$, while a recast of generic multi-lepton searches \cite{CMS:2013jfa}  concluded that and SU(2) singlet $E$ with  $M_E$ in the 100~GeV ballpark is not excluded \cite{Falkowski:2013jya}. 
Furthermore, the mixing angle $\alpha_L$ is constrained by electroweak precision tests. 
At the second-order in $v$ the couplings of the SM  left-handed charged leptons to W and Z are modified as 
\beq
\cL =  \left (1 -  {\alpha_L^2 \over 2} \right ){g_L \over \sqrt 2}  W_\mu^+ \bar \nu _L \gamma_\mu \ell_L
+ \left (  {- g_L^2 + g_Y^2 \over 2 \sqrt{g_L^2 + g_Y^2}}  +     \sqrt{g_L^2 + g_Y^2} {\alpha_L^2 \over 2}  \right )  Z_\mu \bar \ell_L \gamma_\mu \ell_L  . 
\eeq  
The precise constraint on $\alpha_L$ somewhat depends on whether $E$ mixes with $e$, $\mu$, or $\tau$. 
Using the electroweak precision measurements  from LEP-1 and SLC \cite{ALEPH:2005ab} and the recent W mass  measurements  \cite{Group:2012gb} we find   the following 95\% CL limits: 
\bea
\label{eq:alphaLcon}
(e)  & \qquad &  \alpha_L  < 0.017,
\nn
(\mu)  & \qquad &  \alpha_L  < 0.030, 
\nn
(\tau)  & \qquad &  \alpha_L  < 0.050 .
\eea 
For a given  $M_E$ this translates into upper limits on the Yukawa coupling $Y$, and in consequence into upper limits on ${\rm Br}(h \to E \ell)$. 
The maximum allowed branching fractions in the electron, muon and tau channels are  shown in the left panel of \fref{brhEl}.  
These limits turn out to be weak enough to allow an observable signal in the golden channel. 
In fact, the limits on additional width in the golden channel  in \eref{gammalimits} already exclude a sizable chunk of otherwise viable parameter space. 
We conclude that vector-like leptons with mass $M_E \lesssim 125$~GeV can be meaningfully probed by exotic Higgs decays.  

\begin{figure}[!h]
    \begin{center}
          \includegraphics[width=0.55 \textwidth]{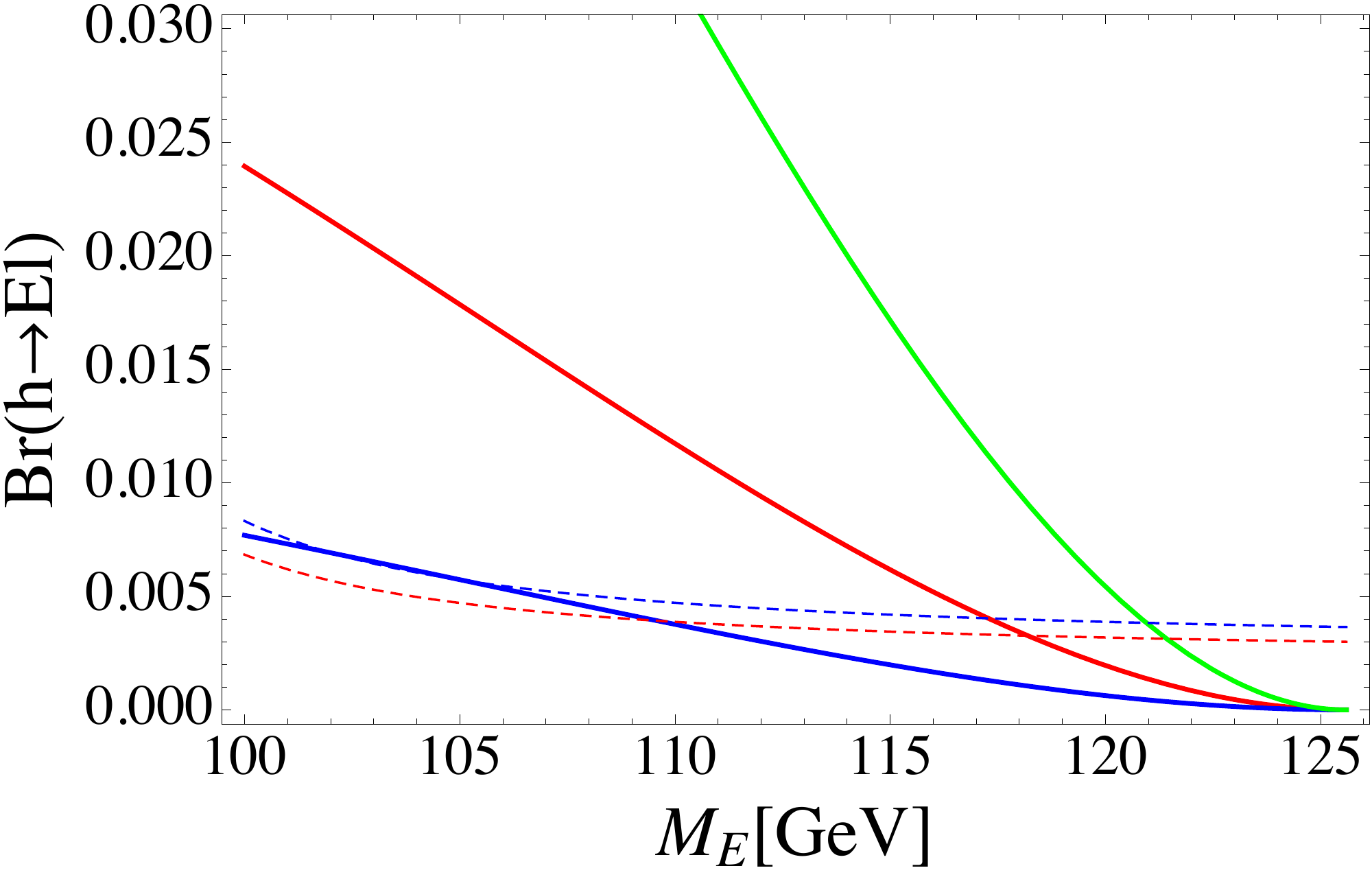}
     \quad 
      \includegraphics[width=0.4\textwidth]{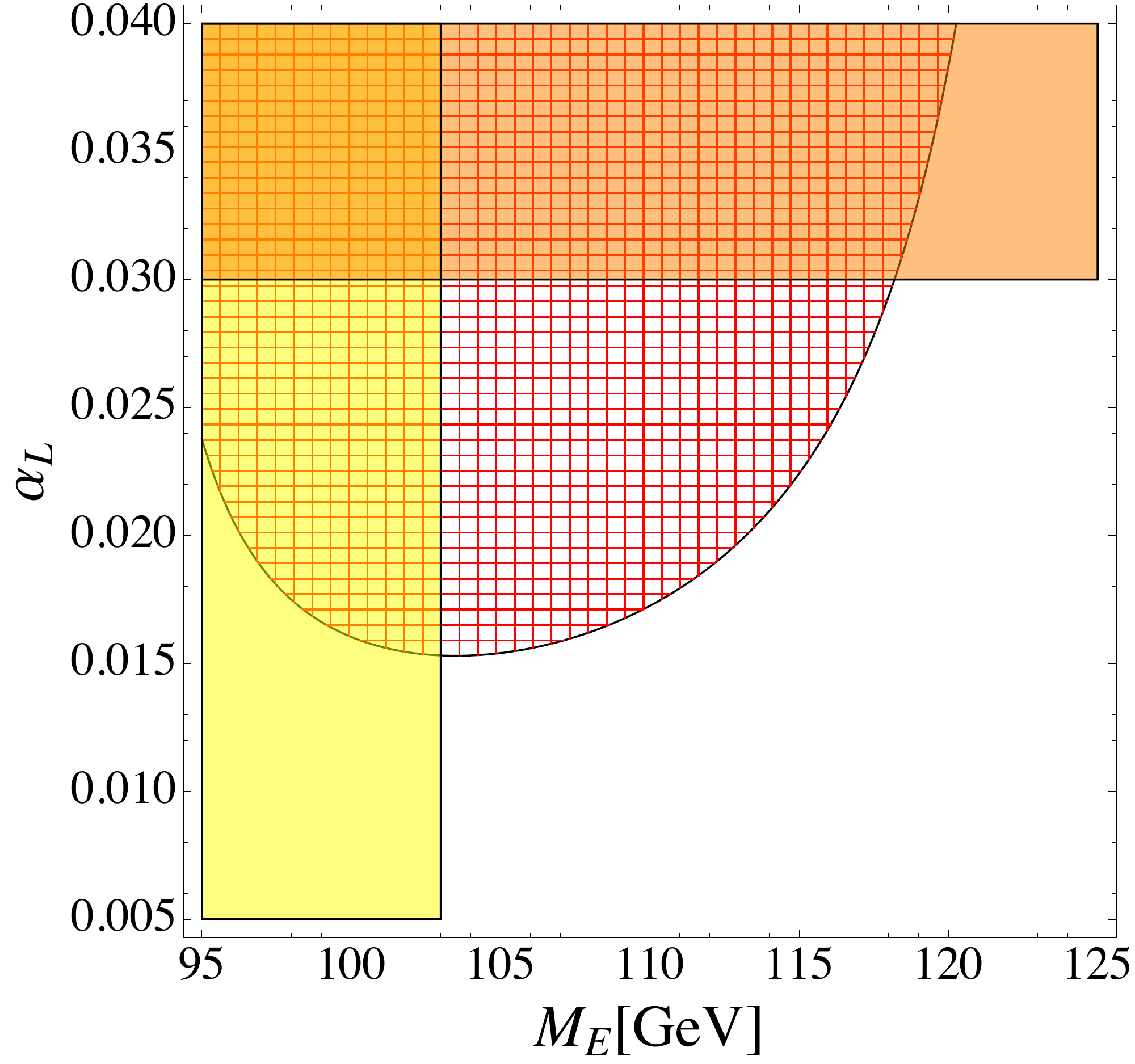}
 \vspace*{-2mm}
          \caption{\footnotesize 
     {\bf Left :}  The maximum branching fraction for $h \to E \ell$ decays allowed by electroweak precision constraints for $\ell =e$ (blue),  $\ell =\mu$ (red), and $\ell =\tau$ (green), as a function of the $E$ mass. The dashed lines indicate the current upper limits on ${\rm Br} (h \to E \ell)$ from the observed $h \to 4$~lepton event rate for $\ell =e$ (blue), and  $\ell =\mu$ (red).               
    {\bf   Right:}   The allowed parameter space in the mass-mixing angle plane for a vector-like SU(2) singlet fermion $E$ mixing with the SM muon. 
       The yellow and orange areas are excluded respectively by direct LEP-2 searches and by electroweak precision constraints.
       The red mesh area is excluded by the observed $h \to 4 \ell$ event rate  (assuming the Higgs couplings to the SM are not modified).       
 }
\label{fig:brhEl}
\end{center}
\vspace*{-3mm}
\end{figure}

\section{Methods} \label{sec:meth}

We are interested in estimating the potential of LHC Higgs searches in the 4-lepton final state to constrain or discover exotic Higgs decays in the models described in \sref{models}.
To distinguish the SM $h \to ZZ^* \to 4 \ell$ decays from those involving a new hidden photon or heavy fermion, we employ a simplified likelihood analysis following closely the procedure used in Ref.~\cite{Stolarski:2012ps} and described in more detail in~\cite{Gao:2010qx,DeRujula:2010ys}.
The $h\rightarrow 4\ell$ channel has a good signal-to-background ratio in the signal region $m_{4 \ell} \approx m_h$, and is very well discriminated from the backgrounds due to the different shapes in the distributions of the various observables~\cite{Chen:2012jy}.
Of course, ideally one would include the dominant $q\bar{q}\rightarrow 4\ell$ background as well in the discriminator in order to make a precise statement about the sensitivity.
However, recent studies~\cite{Chen:2012jy,Chen:2013ejz,Chen:2014pia} indicate that the effects of including the background should be small enough that for the present purposes considering the signal only is sufficient.
  
The starting point for our analysis is an analytic expression for the fully differential $h\rightarrow 2e2\mu$ decay width. 
In the models we consider the decay amplitude receive interfering contributions from the  $h\rightarrow ZZ^* \to 2e2\mu$ diagram and from diagrams with an intermediate  hidden photon or  a vector-like charged fermion.
We use it to build the probability density function  (\emph{pdf}) 
\begin{eqnarray}
\label{eqn:SandB_pdf}
\mathcal{P}_{S}(m_h^2,M_1,M_2,\vec{\Omega} | \vec{\lambda}) &=& \frac{d\Gamma_{h\rightarrow 4\ell}}{dM_1^2dM_2^2d\vec{\Omega}}. 
\end{eqnarray}
Here $M_1$, $M_2$ are the invariant masses of the opposite-sign same-flavor lepton pairs,  and the decay angles $\vec{\Omega} = (\Theta, \cos\theta_1, \cos\theta_2, \Phi_1, \Phi)$  are defined in~\cite{Chen:2014pia}. The $\vec{\lambda}$ represent the parameters of the models to be considered.
To compute the matrix element in  the hidden photon model we modify the results of~\cite{Chen:2012jy}  to include the new gauge boson contribution.
The matrix element in the vector-like lepton model is computed in the  FeynArts/FormCalc framework~\cite{Hahn:2000kx} using a custom model exported from Feynrules~\cite{Christensen:2008py}.
In all cases the interference between the new physics process and the SM is included. 
Throughout we fix the Higgs boson mass as  $m_h = 125.6~$GeV. 

With the \emph{pdfs} at hand we can write the likelihood of obtaining a particular data set containing $N$ events as,
\begin{eqnarray}
\label{eqn:likelihood}
&& L(\vec{\lambda}) = \prod_{\mathcal{O}}^N \mathcal{P}_{S}(\mathcal{O} | \vec{\lambda}), 
\end{eqnarray}
where $\mathcal{O} = (m_h^2,M_1,M_2,\vec{\Omega})$.
We then  construct a simple hypothesis test~\cite{Cousins:2005pq} where as our test statistic we use the log likelihood ratio defined as,
\begin{eqnarray}
\label{eqn:LLhood}
\Lambda = 2\log[{\cal L}(\vec{\lambda}_1)/{\cal L}(\vec{\lambda}_2)] .
\end{eqnarray}
To estimate the expected significance of discriminating between two different hypotheses, we take one hypothesis as true, say $\vec{\lambda}_1$ and generate a set of $N$ $\vec{\lambda}_1$ events. 
We then construct $\Lambda$ for a large number of pseudo-experiments each containing $N$ events in order to obtain a distribution for $\Lambda$.
We repeat this exercise taking $\vec{\lambda}_2$ to be true and obtain a different distribution for $\Lambda$.
With the two distributions for $\Lambda$ in hand we can compute an approximate significance by denoting the distribution with negative mean as $f$ and the distribution with positive mean as $g$ and finding a value  $\hat \Lambda$ such that 
\begin{equation}
\int_{\hat \Lambda}^\infty fdx = \int_{-\infty}^{\hat \Lambda} gdx.
\label{eq:lambdahat}
\end{equation}
We then interpret this probability as a one sided Gaussian $p$-value, which can be used to compute the expected significance for discriminating between hypotheses (see~\cite{Stolarski:2012ps} for more details).
For a simple hypothesis test, this Gaussian approximation is often sufficient~\cite{Cousins:2005pq}.
This procedure is repeated many times for a range of numbers of events $N$ to obtain a significance as a function of $N$ for each hypothesis.
In our simplified framework we have also neglected any detector or production effects, but these effects are small and are not needed for the level of precision we aim for in this study~\cite{Chen:2013ejz,Chen:2014pia}.

For the particular models considered here, $\vec{\lambda}$ corresponds to the mass of the new particle and the model parameters determining their coupling to the Higgs and leptons.   
Specifically, for the hidden photon model $\vec{\lambda} = (m_X, \eps,\eps_2,\eps_3)$, and for the vector-like lepton model $\vec{\lambda} = (M_E,Y)$.   
Our aim is to estimate whether the golden channel can probe the parameter space of these models that is not excluded by precision tests and direct searches.
Various hypothesis tests to this end are conducted in the following section.

\section{Results} \label{sec:results}

In this section we present our results concerning the sensitivity  of  the golden channel to exotic Higgs decays for the models described in \sref{models}. 
To this end we pick a number of benchmarks point near the boundary of the parameter space region allowed by current constraints.
We employ the matrix element approach described in \sref{meth}, where in our hypothesis tests we always compare our new physics model to the SM. 
For a given number $N$ of events in the $h \to 2 e 2 \mu$ channel we perform $1000-10000$ pseudo-experiments to estimate the discriminating power between the SM and hidden photon mediated Higgs decays. 
We repeat this procedure over a range of $N$ to obtain an estimate for the discriminating power as a function of number of events.
For these pseudo-experiments we use the full available information contained in the differential distribution of the 4-lepton final state except for the total integrated event rate -- we refer to this as {\em shape observables}. 
The motivation for separating the total rate is that it is less robust as a discriminator, as it can be affected by physics that has nothing to do with exotic decays, for example by modification of the effective Higgs coupling to gluons.
We find that the discriminating power between the pure SM and hidden photon hypotheses comes mostly from $M_1$ and $M_2$ distributions, whereas angular variables add some discriminating power only in the extended hidden photon model of \eref{steroid}.    
On the other hand, angular variables are important  for separating the signal from the non-Higgs SM background. 
For a number of benchmark points we also show the results of combining  the shape and the total rate observables. 
To reduce computing time, for large $N$ we simply extrapolate our results obtained at lower $N$ assuming the significance grows as $\sqrt{N}$. 
With these tools, we estimate the number of $h \to 2 e 2 \mu$  events required to exclude our benchmark points at a given confidence level.  
Although we do not perform simulations  in the $h \to 4 \mu$ and $h \to 4 e$ channels  we expect that, after combining all 4-lepton channels,  the sensitivity   will correspond roughly to doubling  the number of $h \to 2 e 2 \mu$  events. 
To translate between the number of events  and the LHC luminosity we assume the $27\%$ efficiency of reconstructing 4-lepton Higgs decays (the one in CMS in  the LHC run-I \cite{Chatrchyan:2013mxa}). 
Thus, for example, 300~fb${}^{-1}$ at 14 TeV LHC corresponds to roughly  $275$ $h \to 2e2\mu$ and $600$ $h\to 4 \ell$ expected events, 
where we take $\sigma (p p \to h)  \approx 56$~pb, and ${\rm Br}(h \to 4 \ell) = 1.3 \times 10^{-4}$~\cite{Heinemeyer:2013tqa}.

\begin{table}[h] 
\begin{center}
\begin{tabular}{|c|c|c|c|c|}
 \hline 
 $m_X$ & $\eps$  & $\eps_2$ &  $\eps_3$ &   $R$
\\ \hline 
10 & 0.02 & 0 & 0 & 1.004 
 \\ \hline 
15 & 0.02  & 0 & 0 & 1.006
 \\ \hline 
20 &  0.02 & 0 & 0 &1.019
\\ \hline 
25&  0.02  &  0 & 0 & 1.031
\\ \hline
30&  0.02  & 0 & 0 & 1.039   
\\ \hline 
\hline
30 & 0.02 & 0.01 & 0  &  1.33
\\ \hline 
30& 0.02 & 0 & 0.015 &  1.20
\\ \hline \hline 
35 &  0.02 & 0 & 0 & 1.019
\\ \hline 
40 &  0.02 & 0 & 0 & 1.019
\\ \hline 
50 &  0.02 & 0 & 0 & 1.016
\\ \hline 
60 &  0.018 & 0 & 0 & 1.014
\\ \hline 
\end{tabular}
\qquad \qquad 
\begin{tabular}{|c|c|c|c|c|}
 \hline 
$m_E$ & $\alpha_L$ &   $R$
\\ \hline 
103 & 0.015  &  1.48
\\ \hline
110 & 0.017  &  1.57
\\ \hline
115 & 0.02  &  1.08 
\\ \hline
120 & 0.02  &  0.95 
\\ \hline
\end{tabular}
\end{center}
\caption{
{\em Left:} benchmarks  point for the hidden photon model. 
The 4-lepton event rate relative to the SM one  $R = {\Gamma (h \to 4 \ell) \over \Gamma(h \to  4 \ell)_{\rm SM} }$ was computed using MadGraph~5 \cite{Alwall:2011uj} after imposing the standard CMS cuts: $p_{T,\ell} > 10$~GeV, $|\eta_\ell| <  2.5$,  and $M_1> 50$~GeV, $M_2> 12$~GeV for opposite-sign, same-flavor lepton pairs. For the $m_X=10$~GeV benchmark a weaker cut  $M_2> 5$~GeV is used, as the standard one cuts away most of the signal. 
For the benchmarks with non-zero $\eps_2$ or $\eps_3$ the rate includes the contribution of diagrams with an intermediate off-shell photon. 
{\em Right:} the same for the vector-like lepton mixing with the SM muon. 
}
\label{tab:bench}
\end{table}


We start with the vanilla version of the hidden photon model that corresponds to setting  $\eps_2 = \eps_3 =0$ in \eref{steroid}.\footnote{See Refs.~\cite{Gopalakrishna:2008dv,Davoudiasl:2012ag,Davoudiasl:2013aya} for previous studies of the LHC sensitivity in this model.}
We fix $\eps = 10^{-2}$ for all benchmarks and consider several values of the hidden photon masses in the range 10-60~GeV.
The benchmark points we studied are summarized in \tref{bench} and our results concerning the LHC sensitivity are shown in \fref{vanillazprime}. 
It is worth noting that for these points the total $h \to 4 \ell$ rate is enhanced merely by a  few percent compared to the SM. 
As this is within the uncertainty on the SM Higgs production cross section, the total rate information is not useful to discriminate between the SM and new physics in this case. 
Nevertheless, taking advantage of the full kinematic information contained in the 4-lepton event leads to a good sensitivity to new physics.   
We find that the parameter space of the hidden photon model allowed by electroweak precision observables can be probed already in the coming Run-II of the LHC. 
In particular, assuming 300~${\rm fb}^{-1}$ at 14~TeV will be collected,  $m_X$ in the range 15-65~GeV  can be probed for $\eps$ near the boundary of the region allowed by precision observables. 
Further increase in sensitivity can be obtained in the high-luminosity phase of the LHC (assuming 3000~${\rm fb}^{-1}$ at 14 TeV) or in the future 100~TeV collider.  
In particular, the reach can be extended\footnote{%
Assuming that the cut on the lepton pair invariant mass can be lowered from the current standard value of 12~GeV.}  down to $m_X = 10$~GeV, below which the strong bounds on the kinetic mixing from B-factories make it difficult to probe the simplest hidden photon model in high-energy colliders.  
Note that the case with $m_X+m_Z > m_h$, where the strictly 2-body decay $h\to Z X$ is forbidden, can also be probed to some extent. 
In this case, the kinematic suppression due to the Z boson being strongly off-shell is partially offset by the fact that the $h Z X$ coupling increases with $m_X$. 
On the other hand, for $m_X$ approaching $m_Z$ the  electroweak precision bounds on $\eps$ become stronger (that's why for the benchmark point with  $m_X = 60$~GeV we had to choose a slightly smaller value of $\eps$).
For this reason,  in the allowed parameter space,  the new physics corrections in the $h\to 4 \ell$ channel  quickly become unobservable for $m_X \gtrsim 70$~GeV.   
Finally, we estimate the reach in the kinetic mixing parameter: at the most favorable hidden photon mass $m_X \approx 30$~GeV the high-luminosity LHC will be able to exclude $\eps$ down to $0.007$.   
The bottom line is that the LHC is capable of exploring new interesting regions of the parameter space, even in the simplest version of the hidden photon model.

\begin{figure}[!h]
    \begin{center}
      \includegraphics[width=0.43\textwidth]{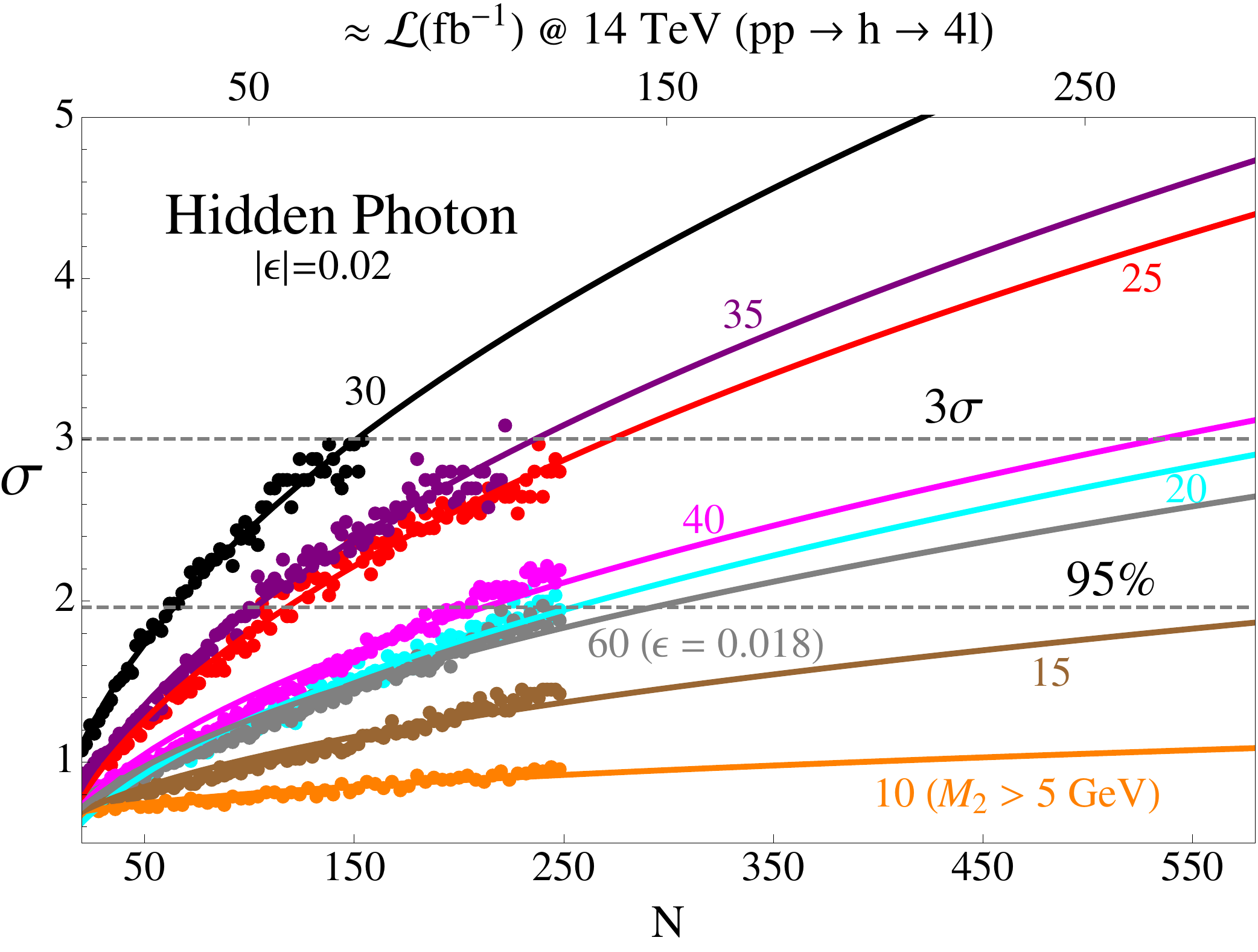}
     \quad 
      \includegraphics[width=0.43\textwidth]{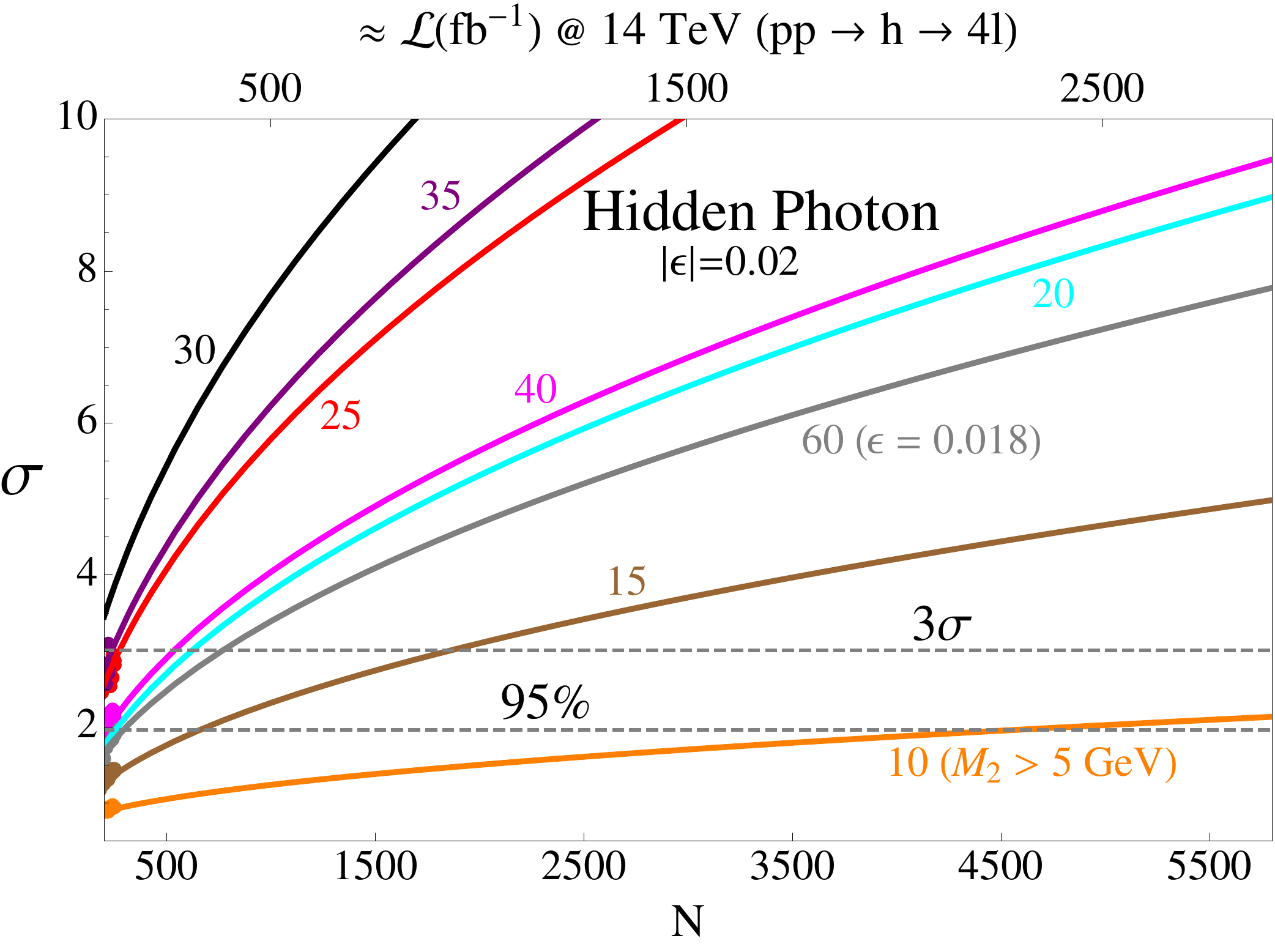}
\vspace*{-2mm}
          \caption{\footnotesize 
       {\bf Left:} The LHC sensitivity for the simplest version of the hidden photon model with $\eps_2 = \eps_3 =0$ and $\eps = 10^{-2}$ for masses ranging from 10~to~60~GeV.
The dots indicate the average $\sigma$ obtain in our set of pseudo experiments which we have conducted for a range of fixed number of events from between $N=20$ and $N=600$.
{\bf Right:} Same, extrapolated to larger $N$, assuming a $\sqrt{N}$ scaling in the sensitivity to estimate the discriminating power at high luminosity.      
 }
\label{fig:vanillazprime}
\end{center}
\vspace*{-3mm}
\end{figure}

The next step is to go beyond the simplest hidden photon model and to allow $\eps_2 \neq 0$ and or $\eps_3 \neq 0$ in \eref{steroid}.
As explained  previously, this extended model allows us to increase new physics corrections to the $h \to 4 \ell$ rate, which greatly improves the sensitivity at the LHC. 
In fact, the strongest constraints on this model are currently provided by the LHC Higgs measurements,  in particular for $m_X=30$~GeV we find $\eps_2 \lesssim 0.015$, $\eps_3 \lesssim 0.02$. 
In the left panel of \fref{exoticzprime1} we show the results for a couple of scenarios with $m_X = 30$~GeV.
Our benchmark points are chosen such that  the  $h \to 4 \ell$ rate is significantly enhanced, by $20$-$30$\%, which is not far from the current upper limit. 
For this reason the rate information alone should be enough to exclude these scenarios at the LHC run-II.  
Taking advantage of the shape information further improves the sensitivity. 
We find that also in this case the shape information has a much stronger discriminating power, as can be clearly seen in the right panel of \fref{exoticzprime1}.
Combining the two, the LHC experiments should be able to comfortably exclude\footnote{Or to discover.} our two benchmarks already after the first year of the coming LHC run.  

 We note that  the discriminating power is increased thanks to the $h X \gamma$ couplings present in the extended model, see \eref{dhgx}.   
 This is partly due to the fact the diagrams with an off-shell photon increase the new physics contribution to the $h \to 4 \ell$ rate.  
But on top off that the  the photon contributions lead to larger shape differences with respect to the SM, primarily in the invariant mass distributions. 
See~\cite{Chen:2014gka} for a study of this effect in a different context. 
Another consequence of the $h X \gamma$  coupling is that the LHC is sensitive to larger values of $m_X$ which would be kinematically suppressed if only $hZX$ couplings were present. This allows the golden channel to probe a larger range of hidden photon masses than might be naively expected, even up to $m_X \sim 100$~GeV.   Finally, we point out  that the golden channel is sensitive not only to the magnitude but also to the signs of $\eps_{2}$ and $\eps_{3}$ relative to that of $\eps$.
Indeed, we find that for the parameter space regions where there is sensitivity to exotic Higgs decays we can discriminate between the positive and negative $\eps_2$ or $\eps_3$ hypotheses. 

\begin{figure}[!h]
    \begin{center}
      \includegraphics[width=0.45\textwidth]{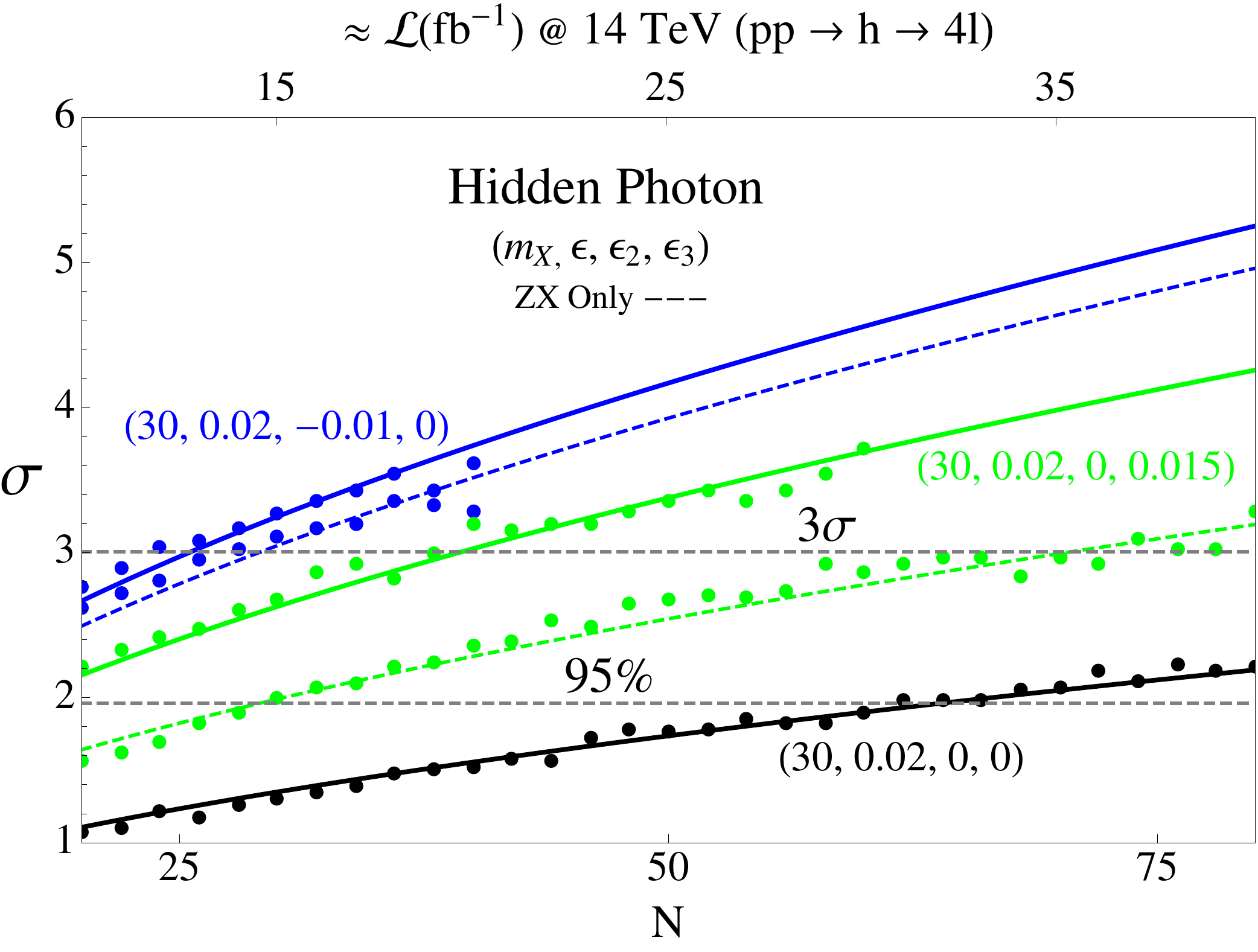}
      \quad
            \includegraphics[width=0.45\textwidth]{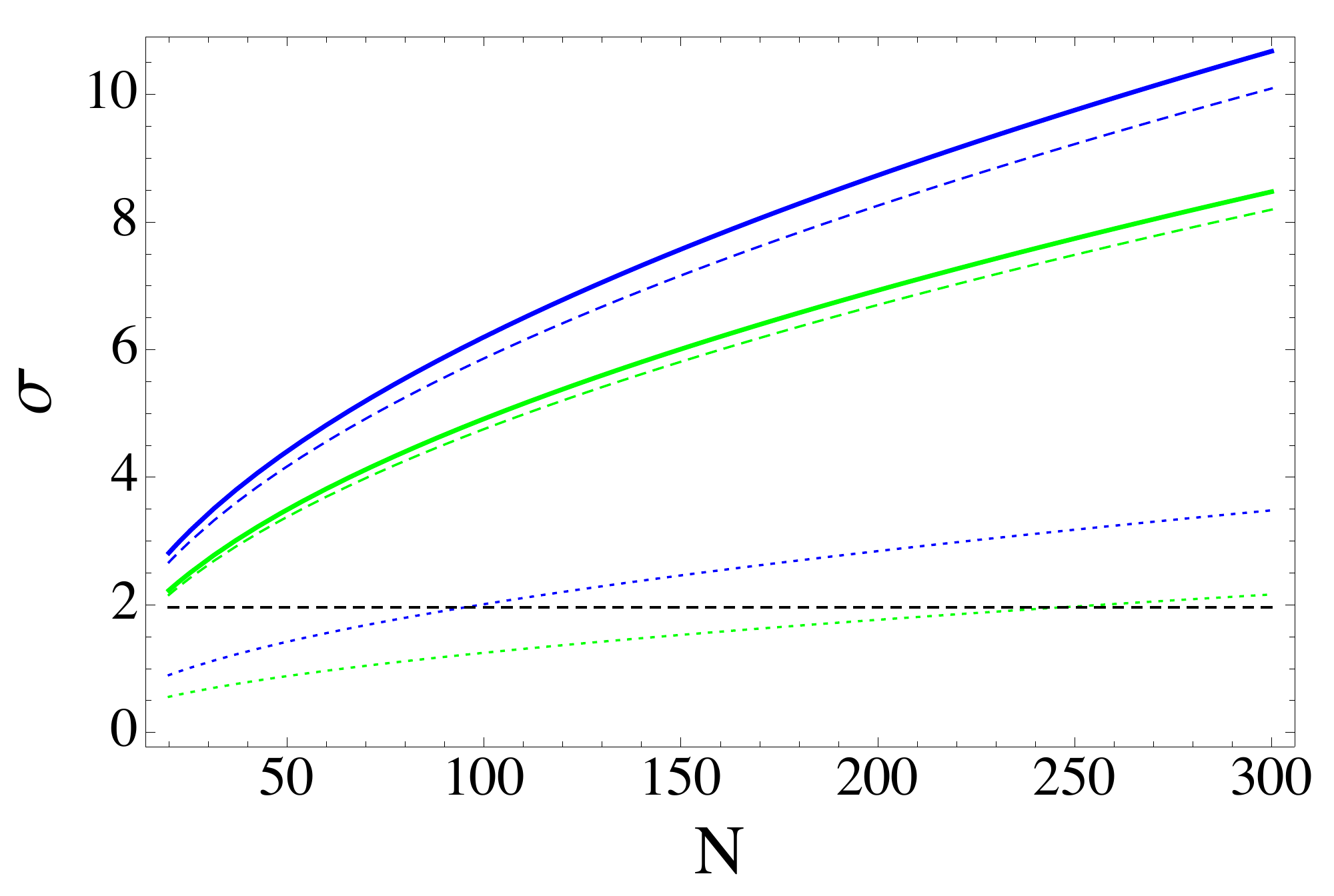}
\vspace*{-2mm}
\caption{\footnotesize 
   {\bf Left:}  LHC sensitivity using the shape of the  4-lepton distribution alone for the extended hidden photon points labeled by the values of $(m_X, \epsilon, \epsilon_2, \epsilon_3)$.
 The dots indicate the results obtained from conducting pseudo experiments which are then extrapolated to larger $N$ assuming the significance grows as $\sqrt{N}$.
 The dashed curves  indicate the sensitivity  when only  the $hXZ$ couplings are taken into account;  the difference between the dashed and solid curves demonstrates the importance of the off-shell photon contributions.
    {\bf Right:} Comparison of the discrimination power using the shape (dashed), rate (dotted), and combined shape+rate information (solid) for the extended hidden photon benchmarks with $m_X = 30$~GeV, $\epsilon=0.02$, and $(\epsilon_2,\epsilon_3) = (0.01,0)$ (blue) and $(\epsilon_2,\epsilon_3) = (0,0.015)$.
 }
\label{fig:exoticzprime1}
\end{center}
\vspace*{-3mm}
\end{figure}

The final exotic Higgs scenario we study here is the vector-like lepton mixing with the SM muon.
The  benchmarks points we analyzed are summarized in \tref{bench}, and the results are shown  in \fref{exoticfermion}. 
We find that in this case the LHC sensitivity is much weaker than in the hidden photon case if only the shape observables are used, see the left panel of \fref{exoticfermion}. 
We also see that the sensitivity quickly decreases as the mass $M_E$ approaches the Higgs boson mass. 
One reason is that ${\rm Br}(h \to E \mu)$ gets kinematically suppressed for $M_E \approx m_h$. 
On top of that, the muon emitted in  the $h \to E \mu$ decay is very soft, therefore it often does not pass experimental cuts. 
Finally, the differential spectrum is much more similar to the SM case than in the hidden photon model. 
All in all, discriminating the vector-like lepton model using shape observables and standard CMS cuts is possible only when large statistics is accumulated, and only in the narrow mass window 103~GeV~$\leq m_E \lesssim$~115~GeV.   
The sensitivity may be improved though by applying additional cuts that target this specific model.  
In particular, the invariant mass of the 3 leptons coming from $E$ decay should reconstruct to $M_E$. 
The combinatorial background can be reduced by constructing $m_{3 \ell}$ out of the 3 hardest leptons in the event, since the muon from $h \to E \mu$ decay is typically soft. 
On the other hand, the total event rate is in this case a much stronger discriminator, as shown in the right panel of  \fref{exoticfermion}.  
Thus,  by simply counting the number of events  in the $2 e 2 \mu$ and $4 \mu$ channels, we can  explore new regions of the $M_E$-$\alpha_L$ parameter space for 103~GeV~$\leq m_E \lesssim 115$~GeV.  
In particular, for $m_E = 103$~GeV we estimate the LHC experiments can probe $\alpha_L$ down to $\sim 0.007$.
Observing an excess of $4\mu$ and  $2 e 2\mu$ events would be a motivation to apply model-specific cuts, to isolate the vector-like lepton signal. 
Similar comments apply to a vector-like lepton   mixing with the SM electron, except that then an excess is expected in the  $4e$ and  $2 e 2\mu$ channels. 
Finally, we note $E$ could mix predominantly with the $\tau$ lepton, which is in fact the most natural possibility from the point of view of models where vector-like leptons play a role in generating the SM fermion mass hierarchies. 
Thus, exploring also the $2 \ell 2 \tau$ final state would be advantageous in this context.

\begin{figure}[!h]
    \begin{center}
      \includegraphics[width=0.45\textwidth]{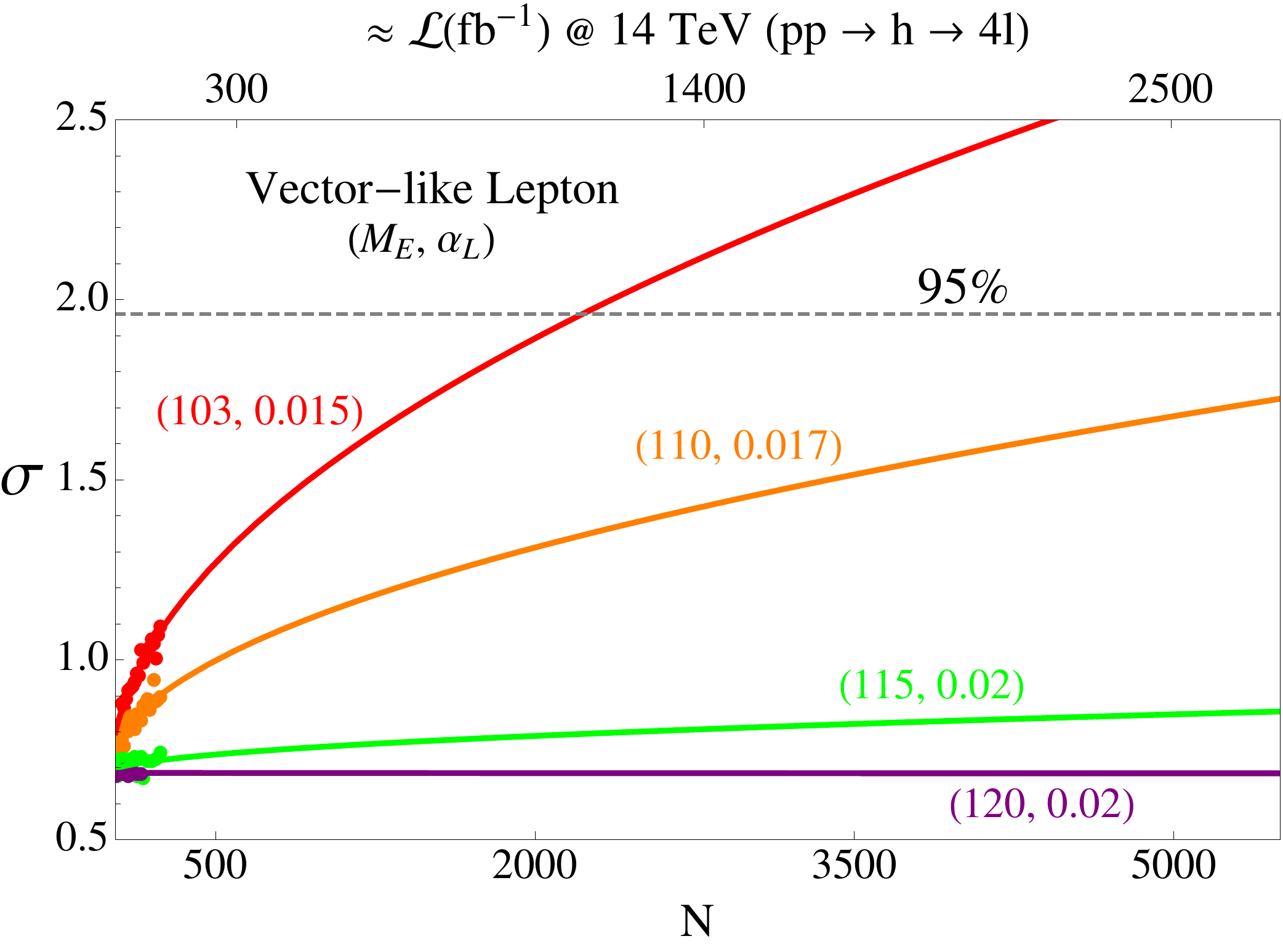}
      \quad 
                  \includegraphics[width=0.45\textwidth]{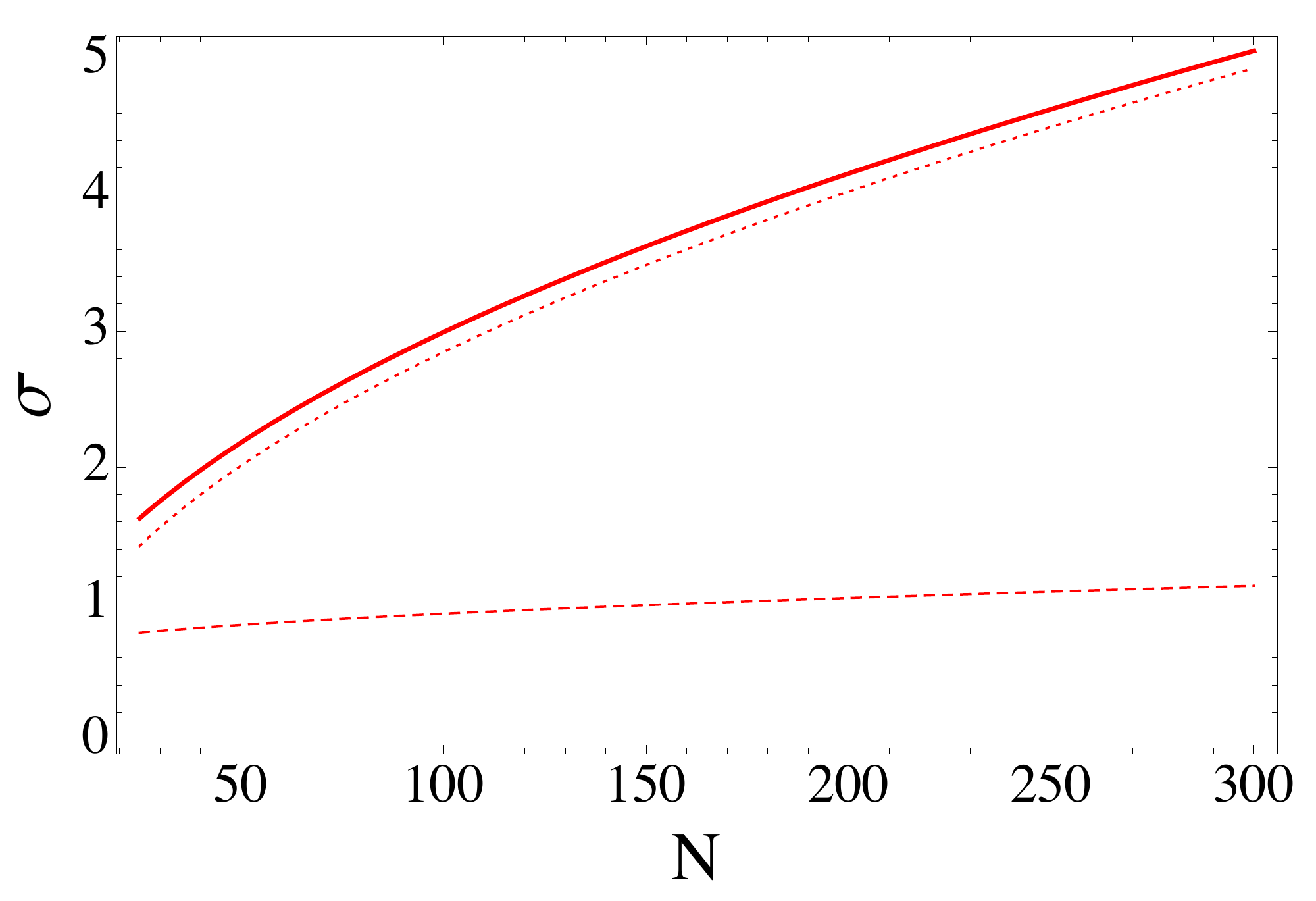}
\vspace*{-2mm}
          \caption{\footnotesize 
   {\bf Left:} LHC sensitivity using the shape of the  4-lepton distribution alone for the vector-like lepton points labeled by the values of $(M_E, \alpha_L)$. 
   The dots indicate the results obtained by conducting pseudo-experiments which are then extrapolated to larger $N$ assuming the significance grows as $\sqrt{N}$.
  {\bf Right:} Comparison of the discrimination power using the shape (dashed), rate (dotted), and combined shape+rate information (solid) for the benchmark point  with $M_E = 103$~GeV, $\alpha_L=0.015$.         
 }
\label{fig:exoticfermion}
\end{center}
\vspace*{-3mm}
\end{figure}

\section{Summary} \label{sec:conclusions}

In this paper we studied the prospects of constraining exotic Higgs decays using the 4-lepton final state.  
We picked two scenarios  of more general interest: a hidden photon mixing with the SM via the hypercharge portal,  and a vector-like charged lepton mixing with  one of the SM leptons via Yukawa interactions.   
Using the rate information only, the LHC run-II is sensitive to exotic decays if the new contributions to the total  $h \to 4 \ell$ rate are larger than $10\%$ of the SM rate. 
This is possible to arrange in the vector-like lepton scenario, and also in the non-minimal hidden photon scenario in the presence of direct Higgs interactions with the hidden sector. 
The main point of this paper is to argue that taking advantage of the full information contained in the differential distribution of the 4-lepton final state dramatically improves the LHC sensitivity. 
To extract that information, we employed the matrix element methods previously developed in the context of measuring the coupling strength and the  tensor structure of Higgs interactions with the SM gauge fields. 
These methods can be carried over to our case  in a straightforward way, as  exotic Higgs decays may readily affect the shape of the 4-lepton differential distribution. 
The shape information is essential in constraining the minimal version of the hidden photon model, where corrections to the total $h \to 4 \ell$ are not expected to exceed a few percent. 
We find that for the hidden photon masses between 15~and~65~GeV the run-II of the LHC will be able to probe a new parameter space of the hidden photon model that is currently allowed by all precision constraints.
Likewise,  in the  non-minimal hidden photon scenario, the shape information allows one to significantly improve the sensitivity such that large chunks of the allowed parameter space can be explored already in the first year of the upcoming LHC run.


\section*{Acknowledgements}

We thank Yi Chen and Kunal Kumar for the help with simulations and validation as well as Jessie Shelton for interesting discussions. 
We also thank Centro de Ciencias de Benasque Pedro Pascual for their hospitality during which much of this work was completed.  
This work was supported by the ERC advanced grant Higgs@LHC.

\bibliographystyle{JHEP}
\bibliography{hdecay}

\end{document}